\documentclass[twocolumn]{aastex701}

\shortauthors{Wei et al.}
\graphicspath{{./}{figures/}}
\usepackage{amsmath}
\usepackage{threeparttable}
\usepackage{booktabs}
\usepackage{changepage} 

\begin{document}
\title{Coupled Shock Cooling and Radioactive Heating in the Type IIb Supernova SN 2024aecx: An Extended Envelope and Rapid Optical Decline}

\author[0009-0006-3261-6418]{Na Wei}
\affiliation{Institute of Astrophysics, Central China Normal University, Wuhan 430079, China; \url{liuld@ccnu.edu.cn;yuyw@ccnu.edu.cn}}
\affiliation{Laboratory for Compact Object Astrophysics and Astronomical Technology, Central 
China Normal University, Wuhan 430079, China}
\affiliation{Education Research and Application Center, National Astronomical Data Center, Wuhan 430079, China}
\email{wei1003@mails.ccnu.edu.cn}

\author[0009-0000-2423-6825]{Yu-Hao Zhang}
\affiliation{Institute of Astrophysics, Central China Normal University, Wuhan 430079, China; \url{liuld@ccnu.edu.cn;yuyw@ccnu.edu.cn}}
\affiliation{Laboratory for Compact Object Astrophysics and Astronomical Technology, Central 
China Normal University, Wuhan 430079, China}
\affiliation{Education Research and Application Center, National Astronomical Data Center, Wuhan 430079, China}
\email{zhang-yh@mails.ccnu.edu.cn}

\author[0000-0002-8708-0597]{Liang-Duan Liu}
\affiliation{Institute of Astrophysics, Central China Normal University, Wuhan 430079, China; \url{liuld@ccnu.edu.cn;yuyw@ccnu.edu.cn}}
\affiliation{Laboratory for Compact Object Astrophysics and Astronomical Technology, Central 
China Normal University, Wuhan 430079, China}
\affiliation{Education Research and Application Center, National Astronomical Data Center, Wuhan 430079, China}
\email{liuld@ccnu.edu.cn}

\author[0000-0001-8744-3813]{Guang-Lei Wu}
\affiliation{Institute of Astrophysics, Central China Normal University, Wuhan 430079, China; \url{liuld@ccnu.edu.cn;yuyw@ccnu.edu.cn}}
\affiliation{Laboratory for Compact Object Astrophysics and Astronomical Technology, Central 
China Normal University, Wuhan 430079, China}
\affiliation{Education Research and Application Center, National Astronomical Data Center, Wuhan 430079, China}
\email{wuguanglei@mails.ccnu.edu.cn}

\author[0009-0004-9719-272X]{Jing-Yao Li}
\affiliation{Institute of Astrophysics, Central China Normal University, Wuhan 430079, China; \url{liuld@ccnu.edu.cn;yuyw@ccnu.edu.cn}}
\affiliation{Laboratory for Compact Object Astrophysics and Astronomical Technology, Central 
China Normal University, Wuhan 430079, China}
\affiliation{Education Research and Application Center, National Astronomical Data Center, Wuhan 430079, China}
\email{lijy@mails.ccnu.edu.cn}

\author[0000-0002-1067-1911]{Yun-Wei Yu}
\affiliation{Institute of Astrophysics, Central China Normal University, Wuhan 430079, China; \url{liuld@ccnu.edu.cn;yuyw@ccnu.edu.cn}}
\affiliation{Laboratory for Compact Object Astrophysics and Astronomical Technology, Central 
China Normal University, Wuhan 430079, China}
\affiliation{Education Research and Application Center, National Astronomical Data Center, Wuhan 430079, China}
\email{yuyw@ccnu.edu.cn}

\author[0000-0002-4731-9698]{Ning-Chen Sun}
\affiliation{School of Astronomy and Space Science, University of Chinese Academy of Sciences, Beijing 100049, China}
\affiliation{National Astronomical Observatories, Chinese Academy of Sciences, Beijing 100101, China}
\affiliation{Institute for Frontiers in Astronomy and Astrophysics, Beijing Normal University, Beijing, 102206, China}
\email{sunnc@ucas.ac.cn}

\begin{abstract}
SN~2024aecx is a nearby, rapidly evolving stripped-envelope supernova
with a prominent double-peaked ultraviolet--optical light curve. We model
its multiband evolution with an extended version of \texttt{TransFit},
in which the early shock-cooling emission and the subsequent
radioactive heating are treated within a single time-dependent radiative
diffusion calculation. To describe the stratified ejecta expected for a
Type~IIb progenitor, we adopt a compact inner ejecta connected to a
dilute extended outer envelope and fit the outer density slope directly
from the early light curve. The model reproduces the short-lived first
peak, the rise to the radioactive main peak, and the overall multiband
evolution. We infer an effective outer radius of
$R_0=109.6^{+6.6}_{-3.5}\,R_\odot$, an ejecta mass of
$M_{\rm ej}=2.14^{+0.21}_{-0.19}\,M_\odot$, a nickel mass of
$M_{\rm Ni}=0.050\pm0.002\,M_\odot$, and a steep outer density slope of
$n_{\rm out}=13.33^{+0.11}_{-0.12}$. The steep outer profile favors a
low-mass extended envelope, while the low ejecta mass explains the rapid
evolution of the main peak. However, a control model with standard
$\gamma$-ray leakage fades too slowly after maximum. We therefore
introduce an effective optical-output factor to quantify the additional
late-time suppression of the ultraviolet--optical luminosity. These
results support the shock-cooling plus radioactive-heating interpretation
of SN~2024aecx, but show that its rapid optical decline requires physics
beyond the simplest radioactive-diffusion prescription.
\end{abstract}


\section{Introduction}
\label{sec:intro}

Type~IIb supernovae (SNe) are stripped-envelope explosions whose
progenitors have lost most, but not all, of their hydrogen envelopes
before core collapse. They show hydrogen features at early phases and
later evolve toward hydrogen-poor stripped-envelope SNe
\citep{Filippenko1988}. Because only a small amount of hydrogen-rich
material remains at the time of explosion, their early light curves are
sensitive probes of the final envelope structure and the pre-supernova
mass-loss history \citep[e.g.,][]{Claeys2011,Lyman2016,
Taddia2018}.

A common photometric signature of Type~IIb SNe is a double-peaked
ultraviolet--optical light curve. The first peak is usually attributed
to shock-cooling emission from an extended, low-mass outer envelope
after shock breakout \citep{Bersten2012,Nakar2014,Piro2015,Sapir2017,
Dessart2018}, whereas the second peak is mainly powered by radioactive
heating from the decay chain
${}^{56}{\rm Ni}\rightarrow{}^{56}{\rm Co}\rightarrow{}^{56}{\rm Fe}$
\citep{Arnett1980,Arnett1982}. The early cooling peak therefore carries
information about the radius, mass, and density structure of the
residual envelope, while the radioactive main peak constrains the ejecta
mass, nickel mass, diffusion timescale, and radioactive-energy
deposition. In many applications, these two phases are modeled as
separate additive components. Such treatments are useful, but they do
not follow the diffusion of shock-deposited and radioactive energy
through the same ejecta structure, and they often describe the extended
material only through an effective radius, envelope mass, or simplified
density gradient.

SN~2024aecx is a nearby, rapidly evolving Type~IIb event in NGC~3521.
It exhibits weak early hydrogen features, a prominent double-peaked
light curve, and a rapid optical decline after the secondary peak
\citep{Zou2026,Xi2026}. Previous studies suggested that its early peak
can be explained by shock cooling from an extended, low-mass envelope,
followed by a radioactively powered main peak \citep{Zou2026,Xi2026}.
However, SN~2024aecx combines a relatively luminous secondary peak with
fast multiband fading after maximum. This behavior makes it a useful
case for testing whether the standard double-peaked Type~IIb picture can
simultaneously explain three connected phases: the short-lived early
cooling peak, the radioactive main peak, and the rapid post-peak optical
decline.

In this work, we model the ultraviolet--optical light curves of
SN~2024aecx using an extended \texttt{TransFit} framework based on
time-dependent radiative diffusion \citep{Liu2025}. The shock-cooling
emission from the initially deposited thermal energy and the subsequent
heating from ${}^{56}{\rm Ni}$ decay are treated within a single
diffusion calculation, rather than as independent additive components.
To describe the stratified ejecta expected for a strongly stripped
Type~IIb progenitor, we adopt a density structure motivated by the
standard picture in which a compact helium-rich core is surrounded by a
low-mass, extended hydrogen-rich envelope
\citep[e.g.,][]{Nomoto1993,Shigeyama1994,Woosley1994,Chevalier2010}.
This Type~IIb-motivated structure allows the early multiband light curve
to constrain the outer-envelope density gradient, rather than only an
effective progenitor radius or envelope mass.

We further test whether the rapid post-peak decline of SN~2024aecx can
be reproduced by radioactive diffusion with standard $\gamma$-ray
leakage. We find that this prescription declines too slowly when the
same physical parameters are constrained by the early cooling peak and
the radioactive main peak. We therefore introduce a late-time effective
optical-output factor to quantify the additional suppression of the
observable optical luminosity. This factor is not intended to identify a
unique microscopic mechanism, but to diagnose missing late-time physics
such as enhanced nonlocal energy escape, redistribution of optical
radiation to the infrared, dust effects, ejecta geometry, or opacity
evolution. In this framework, the early cooling peak, the radioactive
main peak, and the rapid late-time optical fading of SN~2024aecx are
treated as linked constraints on a single time-dependent diffusion model.

\section{Observational Properties and Empirical Constraints}
\label{sec:obs}

SN~2024aecx is located in the nearby spiral galaxy NGC~3521, with a host-galaxy redshift of $z=0.002665$. 
\citet{Xi2026} derived a distance of $D=11.3\pm1.1~{\rm Mpc}$ using the tip of the red giant branch method. 
The transient was first detected by ATLAS at MJD~60660.56 \citep{Stevance2024}. 
Non-detections from ATLAS and ZTF within about one day before discovery \citep{Perez2024} provide a tight constraint on the explosion epoch. 
We adopt $T_0={\rm MJD}~60659.953$, estimated by \citet{Zou2026} from the midpoint between the last non-detection and the first detection. 

\begin{deluxetable*}{lcccccccc}
\tablecaption{
Empirical light-curve parameters of SN~2024aecx. 
\label{tab:gp_parameters}
}
\tablehead{
\colhead{Filter} &
\colhead{$t_{\rm min}$} &
\colhead{$m_{\rm min}$} &
\colhead{$t_{\rm p}$} &
\colhead{$m_{\rm p}$} &
\colhead{$\Delta m_{15}$} &
\colhead{$t_{\rm rise}$} &
\colhead{$\dot{m}_{\rm early}$} &
\colhead{$\dot{m}_{15}$} \\
\colhead{} &
\colhead{(day)} &
\colhead{(mag)} &
\colhead{(day)} &
\colhead{(mag)} &
\colhead{(mag)} &
\colhead{(day)} &
\colhead{(${\rm mag~d^{-1}}$)} &
\colhead{(${\rm mag~d^{-1}}$)}
}
\startdata
$\rm B$ & 3.57 & $16.78\pm0.06$ & 19.69 & $15.16\pm0.06$ & -- & 16.12 & $0.646\pm0.022$ & -- \\
$\rm V$ & 5.04 & $16.15\pm0.15$ & 19.26 & $14.34\pm0.05$ & -- & 14.22 & $0.281\pm0.036$ & -- \\
$\rm R$ & 4.30 & $15.54\pm0.07$ & 20.49 & $14.09\pm0.04$ & -- & 16.19 & $0.104\pm0.020$ & -- \\
$\rm I$ & 4.41 & $15.62\pm0.07$ & 21.17 & $13.97\pm0.04$ & -- & 16.76 & $0.110\pm0.020$ & -- \\
$g$ & 5.15 & $16.49\pm0.06$ & 18.79 & $14.69\pm0.02$ & $1.67\pm0.03$ & 13.64 & $0.297\pm0.015$ & $0.111\pm0.002$ \\
$r$ & 4.15 & $15.72\pm0.06$ & 20.52 & $14.03\pm0.02$ & $1.27\pm0.03$ & 16.38 & $0.173\pm0.020$ & $0.085\pm0.002$ \\
$i$ & 4.35 & $15.67\pm0.06$ & 20.66 & $14.00\pm0.03$ & $1.14\pm0.05$ & 16.31 & $0.157\pm0.019$ & $0.076\pm0.003$ \\
$z$ & 4.02 & $15.53\pm0.05$ & 21.64 & $13.84\pm0.04$ & $1.02\pm0.04$ & 17.62 & $0.088\pm0.018$ & $0.068\pm0.003$ \\
$\rm o$ & 4.84 & $15.27\pm0.10$ & 20.72 & $13.87\pm0.04$ & $1.32\pm0.08$ & 15.88 & $0.087\pm0.026$ & $0.088\pm0.005$ \\
\enddata
\tablecomments{
The parameters are measured from Gaussian-process interpolated optical light curves. 
$t_{\rm min}$ and $t_{\rm p}$ denote the valley and secondary-peak epochs, respectively. 
$t_{\rm rise}=t_{\rm p}-t_{\rm min}$, and $\dot{m}_{15}=\Delta m_{15}/15~{\rm d}$. 
Post-peak quantities are reported only for bands with sufficient coverage after the secondary peak.
}
\end{deluxetable*}

SN~2024aecx has been discussed as a rapidly evolving Type~IIb event in optical studies \citep{Hinkle2024,Andrews2024a,Xi2026,Zou2026}. 
Its early spectra showed weak hydrogen features that rapidly weakened or disappeared within about $30~{\rm d}$, while its ultraviolet--optical light curves displayed a clear double-peaked morphology. 
A Type~Ic interpretation has also been suggested from early classification and near-infrared spectroscopy \citep{Andrews2024b,Tinyanont2026}. 
In this work, we adopt the Type~IIb double-peaked light-curve framework and focus on the origin of the early cooling peak, the radioactively powered main peak, and the rapid post-peak optical decline.

The ultraviolet--optical light curves show three phases that are directly relevant to the modeling. 
An early peak appears within about one day after explosion and is followed by a rapid decline to a valley. 
The light curves then rise to a brighter secondary peak at about $19$--$22~{\rm d}$ and fade rapidly after maximum in several optical bands. 
This morphology is broadly consistent with shock cooling from an extended outer envelope followed by radioactive heating, but the fast post-peak decline provides an additional constraint beyond the standard double-peaked Type~IIb picture.

To quantify the empirical light-curve properties, we interpolate each optical-band light curve with Gaussian Process Regression and measure the valley epoch, the secondary-peak epoch, the rise time, and the post-peak decline rate. 
The valley epoch is denoted by $t_{\rm min}$, and the secondary-peak epoch by $t_{\rm p}$. 
The rise time from the valley to the secondary peak is $t_{\rm rise}=t_{\rm p}-t_{\rm min}.$

The post-peak decline amplitude is defined as
\begin{equation}
\Delta m_{15}=m(t_{\rm p}+15~{\rm d})-m(t_{\rm p}),
\end{equation}
with the corresponding average decline rate $\dot{m}_{15}=\Delta m_{15}/{15~{\rm d}}$. 
The early decline rate, $\dot{m}_{\rm early}$, is measured between the first peak and the valley.

The resulting empirical parameters are summarized in Table~\ref{tab:gp_parameters}. 
For the optical bands with adequate coverage around the valley and the secondary peak, we find $t_{\rm p}\simeq19$--$22~{\rm d}$ and $t_{\rm rise}\simeq14$--$18~{\rm d}$. 
The decline after the first peak is wavelength dependent, with faster fading in the bluer bands and slower evolution toward redder bands. 
After the secondary peak, SN~2024aecx declines rapidly in the well-sampled $g,r,i,o,z$ bands, with $\Delta m_{15}\simeq1.02$--$1.67~{\rm mag}$ and $\dot{m}_{15}\simeq0.068$--$0.111~{\rm mag~d^{-1}}$. 
The Swift/UVOT ultraviolet bands and the ATLAS $c$ band are not used to define the optical secondary-peak parameters because of limited pre-peak coverage, but they are included as auxiliary constraints in the full multiband modeling.

The observed light curves therefore impose three direct requirements on the model. 
First, the short-lived early peak and its wavelength-dependent decline require a treatment sensitive to the radius and density structure of the outermost ejecta. 
Second, the timing and luminosity of the secondary peak require radioactive heating coupled to the ejecta diffusion timescale. 
Third, the rapid decline after the secondary peak must be reproduced across multiple optical bands, providing a direct test of whether standard radioactive diffusion with $\gamma$-ray leakage is sufficient. 
These requirements motivate the unified time-dependent radiative-diffusion framework described below, in which the early shock-cooling emission, the subsequent ${}^{56}{\rm Ni}$ heating, and the additional late-time suppression of the observable optical output are treated within a common modeling framework.

\section{A Unified Time-dependent Diffusion Model}
\label{sec:model}

We model the ultraviolet--optical light curves of SN~2024aecx with an
extended version of the time-dependent radiative-diffusion code
\texttt{TransFit} \citep{Liu2025}. The goal of the model is not to fit
the early and late parts of the light curve with independent components,
but to describe the shock-cooling peak, the radioactive main peak, and
the rapid post-peak optical decline within a single diffusion-based
framework.

The baseline \texttt{TransFit} calculation follows the diffusion of
radiation through homologously expanding ejecta. The initial
shock-deposited thermal energy produces the early cooling emission, while
radioactive heating is included as a time-dependent source term and powers the
secondary peak. For SN~2024aecx, we extend this framework in two ways.
First, we introduce a Type~IIb-motivated three-zone ejecta density
structure, which allows the early multiband light curve to constrain the
outer-envelope density gradient. Second, we introduce an effective
late-time optical-output factor to quantify the additional suppression
of the observed optical luminosity beyond standard radioactive diffusion
with $\gamma$-ray leakage.

\subsection{Baseline Diffusion Framework}
\label{subsec:baseline_diffusion}

We assume homologous expansion and write the outer ejecta radius as
\begin{equation}
R_{\rm out}(t)=R_0+v_{\max}t ,
\end{equation}
where $R_0$ is the initial outer radius and $v_{\max}$ is the maximum
ejecta velocity. In this work, $R_0$ is interpreted as an effective
radius of the progenitor or extended envelope at the onset of the
cooling phase, rather than as a detailed hydrostatic stellar radius. It
sets the initial emitting scale and affects the adiabatic degradation of
the shock-deposited thermal energy before photon escape.

Using the comoving coordinate
\begin{equation}
x\equiv \frac{r}{R_{\rm out}(t)}, \qquad 0\leq x\leq 1 ,
\end{equation}
the ejecta density is written as
\begin{equation}
\rho(r,t)=\rho_0
\left[\frac{R_0}{R_{\rm out}(t)}\right]^3
\eta_{\rm ej}(x),
\end{equation}
where $\eta_{\rm ej}(x)$ is the dimensionless density profile.

The radiation internal-energy density is expressed as
\begin{equation}
u(r,t)=u_0
\left[\frac{R_0}{R_{\rm out}(t)}\right]^4
e(x,t),
\end{equation}
where the factor $[R_0/R_{\rm out}(t)]^4$ accounts for adiabatic dilution
in homologous expansion. The dimensionless energy variable $e(x,t)$
then evolves according to the time-dependent diffusion equation
\begin{equation}
\frac{\partial e}{\partial t}
=
\frac{1}{x^2}
\frac{\partial}{\partial x}
\left[
D(x,t)\frac{\partial e}{\partial x}
\right]
+
S(x,t),
\end{equation}
where the first term describes radiative diffusion and $S(x,t)$ denotes
local heating. In the present model, the early cooling peak is produced
by the diffusion of the initial shock-deposited thermal energy, while
the radioactive contribution enters through $S(x,t)$ and powers the
main peak.

The bolometric luminosity is obtained from the diffusive flux at the
outer boundary,
\begin{equation}
L_{\rm bol}(t)
=
-4\pi R_{\rm out}^2(t)
\left.
\frac{c}{3\kappa\rho}
\frac{\partial u}{\partial r}
\right|_{R_{\rm out}(t)} ,
\end{equation}
where $\kappa$ is the gray optical opacity.

\begin{figure}
    \centering
    \includegraphics[width=1\linewidth]{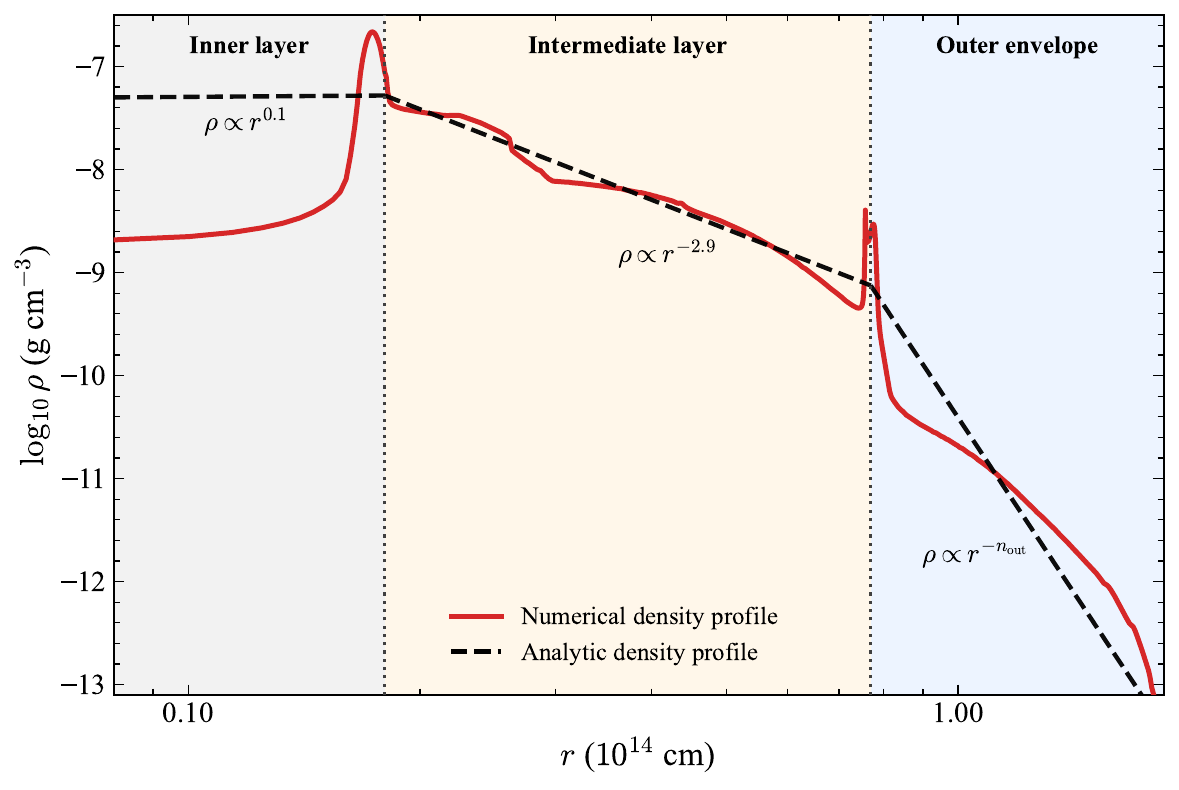}
    \caption{
Density profile of a representative Type~IIb ejecta model at
approximately $t\simeq1~{\rm d}$ after explosion and its continuous
broken power-law approximation. The solid curve shows the density profile obtained from a SNEC simulation \citep{Morozova2015}, while the dashed curve shows the analytic approximation
used in the diffusion calculation. The approximation is calibrated to
capture the global stratified structure of the inner ejecta, the
transition region, and the dilute extended envelope. 
    }
    \label{fig:rho-r}
\end{figure}

\subsection{Numerically Motivated Type~IIb Ejecta Density Profile}
\label{subsec:density}

The early shock-cooling emission is sensitive to the density and
optical-depth structure of the outer ejecta. For extended material after
shock breakout, the relevant structure is the density profile in the
homologous-expansion phase, where the outer layers expand faster and the
early luminosity is generated by the diffusion of shock-deposited
thermal energy through these low-mass layers. Previous analytic and
numerical studies have shown that such extended material can be
approximated by a broken power-law structure with a relatively shallow
inner part and a steep outer tail \citep{Chevalier1989,Piro2021}. This
motivates using the outer density gradient as a key parameter for
describing the early shock-cooling signal.

In this work, we adopt an ejecta density profile motivated by numerical
Type~IIb progenitor models. A Type~IIb progenitor is expected to contain
a compact helium-rich core surrounded by a low-mass, extended
hydrogen-rich envelope \citep[e.g.,][]{Nomoto1993,Shigeyama1994,
Woosley1994,Chevalier2010}. After explosion, this structure does not
produce a single power-law density profile. Instead, the homologously
expanding ejecta shows a stratified structure, with the bulk ejecta
connected to a dilute outer envelope through a transition region. We
therefore use a continuous broken power-law approximation to capture the
global numerical density profile and embed it directly into the
time-dependent diffusion calculation.

The inner and transition parts of the density profile are fixed to match
the representative Type~IIb numerical model shown in
Figure~\ref{fig:rho-r}. The outer density slope, denoted by
$n_{\rm out}$, is treated as a free parameter because the early
multiband light curve is most sensitive to the structure of the dilute
outer envelope. A steeper outer profile places less mass at large radii
and allows the diffusion wave to cross the shock-heated material more
rapidly. This leads to a narrower early cooling peak, a faster decline
toward the valley, and a more rapid early color evolution. A shallower
outer profile retains more optically thick material in the extended
envelope and therefore broadens the cooling phase.

This prescription should be regarded as a compact representation of the
numerical Type~IIb ejecta structure rather than a unique inversion of the
outermost density profile. Numerical models can contain local density
features associated with composition interfaces, while the observed
early light curve mainly probes the density structure averaged over the
layers crossed by the diffusion wave. The purpose of this treatment is
therefore to retain the outer-envelope density information most relevant
for shock cooling while keeping the number of free structural parameters
small.

\subsection{Shock-deposited and Radioactive Heating}
\label{subsec:heating}

The two peaks of SN~2024aecx are powered by different energy reservoirs
but are treated within the same diffusion calculation. The first peak
arises from the cooling diffusion of shock-deposited thermal energy in
the extended outer ejecta. This energy is included in the initial
radiation-energy distribution and subsequently evolves through adiabatic
expansion and radiative diffusion.

The secondary peak is powered mainly by radioactive heating from the
decay chain
${}^{56}{\rm Ni}\rightarrow{}^{56}{\rm Co}\rightarrow{}^{56}{\rm Fe}$.
We adopt the standard radioactive heating rate
\citep{Arnett1982,Nadyozhin1994,Valenti2008}
\begin{equation}
L_{\rm rad}(t)=M_{\rm Ni}
\left[
\epsilon_{\rm Ni}e^{-t/\tau_{\rm Ni}}
+
\epsilon_{\rm Co}
\left(e^{-t/\tau_{\rm Co}}-e^{-t/\tau_{\rm Ni}}\right)
\right],
\end{equation}
where $M_{\rm Ni}$ is the synthesized nickel mass, $\epsilon_{\rm Ni}$
and $\epsilon_{\rm Co}$ are the specific heating rates, and
$\tau_{\rm Ni}$ and $\tau_{\rm Co}$ are the corresponding decay
timescales. The spatial distribution of radioactive heating is
controlled by a radial mixing function and enters the diffusion equation
through the source term $S(x,t)$.

As the ejecta expands, $\gamma$-ray escape reduces the deposited
radioactive power. We write the $\gamma$-ray optical depth as
\begin{equation}
\tau_\gamma(t)=\left(\frac{t_\gamma}{t}\right)^2 ,
\end{equation}
and the deposition fraction as
\begin{equation}
f_\gamma(t)=1-\exp[-\tau_\gamma(t)].
\end{equation}
The deposited radioactive luminosity is then
\begin{equation}
L_{\rm dep}(t)=f_\gamma(t)L_{\rm rad}(t).
\end{equation}
This defines the standard radioactive-diffusion model with
$\gamma$-ray leakage used as the baseline late-time prescription.

\subsection{Effective Late-time Optical-output Factor}
\label{subsec:late_output}

For SN~2024aecx, the standard radioactive-diffusion model with
$\gamma$-ray leakage does not provide a sufficiently rapid post-peak
decline in the observed optical bands. To quantify the additional
suppression of the optical luminosity, we introduce an effective
late-time optical-output factor $f_{\rm out}(t)$, applied to the
baseline diffusion luminosity:
\begin{equation}
L_{\rm opt}(t)=L_{\rm bol}(t)f_{\rm out}(t).
\end{equation}
Here $L_{\rm bol}(t)$ denotes the baseline diffusive luminosity computed
from the radiation-energy equation, and $L_{\rm opt}(t)$ denotes the
effective luminosity emerging in the observed ultraviolet--optical
bands.

Conceptually, we write
\begin{equation}
f_{\rm out}(t)=
\begin{cases}
1, & t\leq t_b,\\
\exp[-(t-t_b)/\tau_{\rm out}], & t>t_b ,
\end{cases}
\end{equation}
where $t_b$ is the onset time of the additional optical suppression and
$\tau_{\rm out}$ is the corresponding fading timescale. In the numerical
implementation, the two branches are connected smoothly to avoid
artificial discontinuities.

The factor $f_{\rm out}(t)$ is not an additional energy source and
should not be interpreted as the microscopic thermalization efficiency
of radioactive decay products. The baseline model already includes
$\gamma$-ray leakage through $f_\gamma(t)$, so $f_{\rm out}(t)$ instead
describes an additional reduction in the effective ultraviolet--optical
output beyond standard radioactive diffusion and $\gamma$-ray escape.
Possible physical contributions include nonlocal energy escape,
redistribution of optical radiation into the infrared, dust formation or
a dust echo, ejecta asphericity or clumping, and changes in opacity or
line-formation processes at late times
\citep{Swartz1995,Maeda2008,Jerkstrand2015,Gall2014,Dimitriadis2017}.
In this work, we use $f_{\rm out}(t)$ as a phenomenological diagnostic
of missing late-time physics required by the rapid post-peak optical
decline.

\section{Light-curve Fitting and Results}
\label{sec:lc_fitting}

\begin{deluxetable*}{lccccccc}
\tabletypesize{\footnotesize}
\setlength{\tabcolsep}{3.5pt}
\tablewidth{0pt}
\tablecaption{Comparison of model parameters for SN~2024aecx.
\label{tab:model_summary}}
\tablehead{
\colhead{Model} &
\colhead{$R_{0}$} &
\colhead{$M_{\rm env}$} &
\colhead{$E_{\rm Th,in}$} &
\colhead{$M_{\rm ej}$} &
\colhead{$M_{\rm Ni}$} &
\colhead{$t_{\rm shift}$} &
\colhead{Ref.} \\
\colhead{} &
\colhead{($R_\odot$)} &
\colhead{($M_\odot$)} &
\colhead{($10^{49}\ {\rm erg}$)} &
\colhead{($M_\odot$)} &
\colhead{($M_\odot$)} &
\colhead{(d)} &
\colhead{}
}
\startdata
Shock cooling, $n=3$
& $169.15^{+1.23}_{-1.27}$
& $0.2400^{+0.0003}_{-0.0004}$
& \nodata
& \nodata
& \nodata
& $0.550^{+0.005}_{-0.005}$
& \citet{Zou2026} \\
Shock cooling, $n=3/2$
& $200.35^{+0.73}_{-0.70}$
& $0.03000^{+0.00003}_{-0.00003}$
& \nodata
& \nodata
& \nodata
& $1.0000^{+0.0002}_{-0.0001}$
& \citet{Zou2026} \\
Two-component model
& $66.75^{+25.47}_{-25.93}$
& $0.04\pm0.01$
& $0.96^{+0.97}_{-0.36}$
& $1.55^{+0.18}_{-0.14}$
& $0.09\pm0.01$
& $0.02^{+0.02}_{-0.01}$
& \citet{Xi2026} \\
\texttt{TransFit}
& $109.59^{+6.61}_{-3.51}$
& \nodata
& $9.64^{+0.27}_{-0.67}$
& $2.14^{+0.21}_{-0.19}$
& $0.050\pm0.002$
& $0.780\pm0.003$
& This work \\
\enddata
\tablecomments{
A dash indicates that the corresponding parameter is not included
or was not reported in that model.
}
\end{deluxetable*}

We fit the multiband ultraviolet--optical light curves of SN~2024aecx with
the extended \texttt{TransFit} model using the Markov Chain Monte Carlo (MCMC) ensemble sampler implemented in \texttt{emcee} \citep{Foreman2013}. We fix $\kappa=0.1~{\rm cm^2~g^{-1}}$,
$\kappa_\gamma=0.03~{\rm cm^2~g^{-1}}$, and adopt
$v_{\rm ej}=6500~{\rm km~s^{-1}}$ based on the Fe~II velocity near maximum light \citep{Xi2026}.

\begin{figure*}
    \centering
    \includegraphics[width=0.8\linewidth]{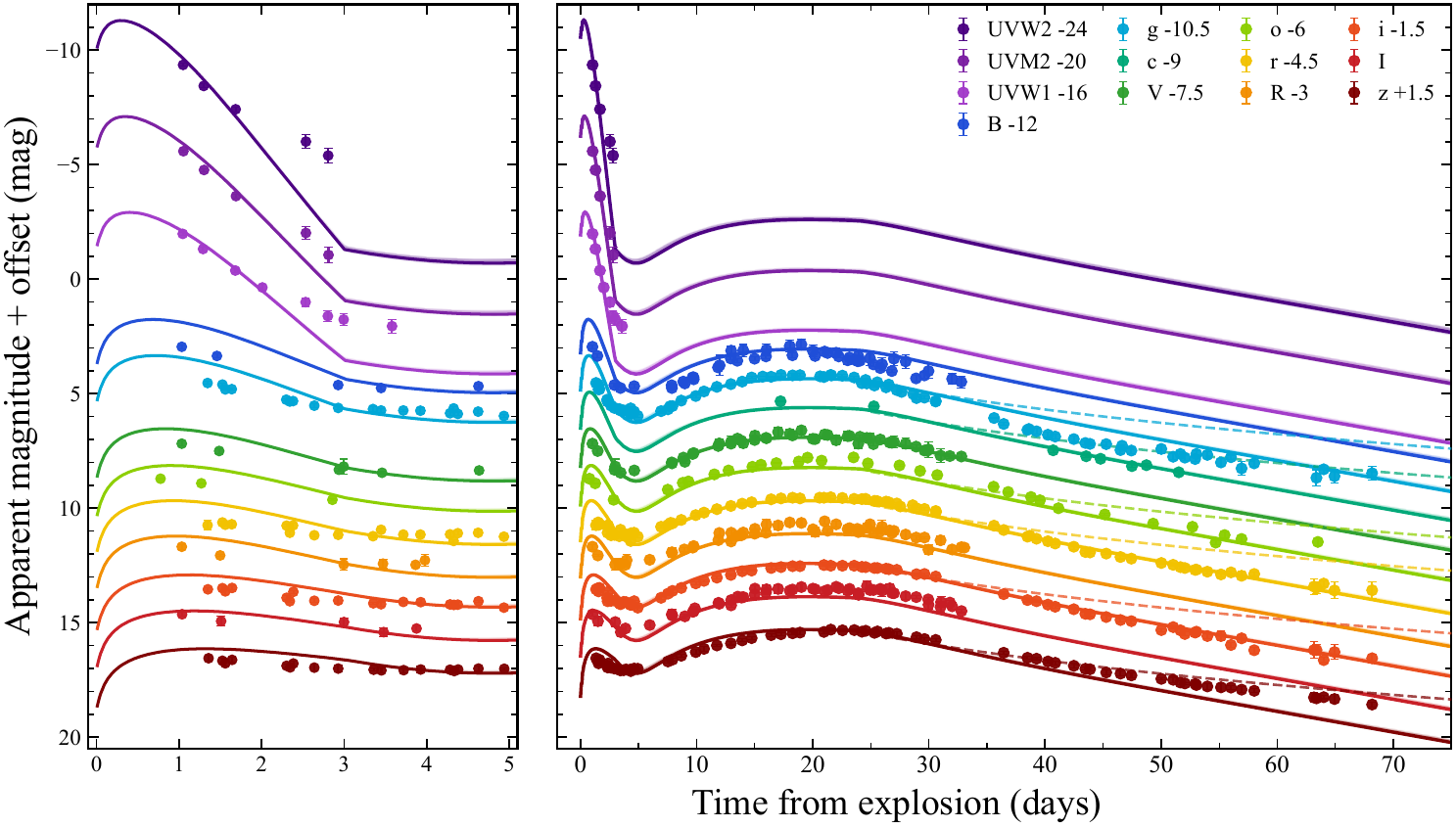}
    \caption{
  Multiband light-curve fit of SN~2024aecx with the extended
    \texttt{TransFit} model. The left panel zooms in on the first
    \(5~{\rm d}\) after explosion, highlighting the early shock-cooling
    phase. The right panel shows the full ultraviolet--optical evolution.
    The circles show the observed photometry, and the solid curves show
    the best-fit model including the late-time optical-output factor
    \(f_{\rm out}(t)\). The dashed curves show the control model with
    \(f_{\rm out}=1\), using the same physical parameters, to illustrate
    the failure of standard radioactive diffusion with \(\gamma\)-ray
    leakage alone to reproduce the rapid post-peak decline. The light
    curves are vertically shifted for clarity, as indicated in the
    legend.
    }
    \label{fig:Multiband}
\end{figure*}

Figure~\ref{fig:Multiband} presents the main multiband fitting result.
The model reproduces the overall ultraviolet--optical evolution of
SN~2024aecx, including the short-lived early peak, the rise to the
radioactive main peak, and the rapid optical fading after maximum. The
left panel zooms in on the first few days after explosion. The early
emission is produced by the cooling diffusion of shock-deposited thermal
energy in the extended outer ejecta. The model captures the main
wavelength-dependent behavior of this phase: the ultraviolet and blue
bands fade more rapidly, while the redder optical bands evolve more
slowly. The detailed morphology of the first peak is not reproduced
point by point, which may reflect simplifications in the outermost
density structure, temperature evolution, composition stratification, or
the blackbody approximation.

The right panel of Figure~\ref{fig:Multiband} shows the full
ultraviolet--optical evolution. The secondary peak is mainly powered by
radioactive heating. The inferred nickel mass is sufficient to account for the
main-peak luminosity, while the relatively low ejecta mass gives a short
effective diffusion timescale. The post-peak decline provides the
strongest test of the model: the control calculation with
\(f_{\rm out}=1\) declines too slowly even with \(\gamma\)-ray leakage,
whereas the full model including \(f_{\rm out}(t)\) follows the rapid
multiband fading more closely.

Compared with previous analyses (see Table~\ref{tab:model_summary}), our inferred effective radius is smaller than the two analytic shock-cooling estimates of \citet{Zou2026} and
is larger than that inferred by the two-component model of
\citet{Xi2026}. The ejecta mass is comparable to, although somewhat
higher than, the value obtained by \citet{Xi2026}. The main difference
of the present model is that the early cooling emission and the
radioactive main peak are computed within the same time-dependent
diffusion calculation, with the outer density slope retained as a
structural parameter.

\begin{figure*}
    \centering
    \includegraphics[width=0.8\linewidth]{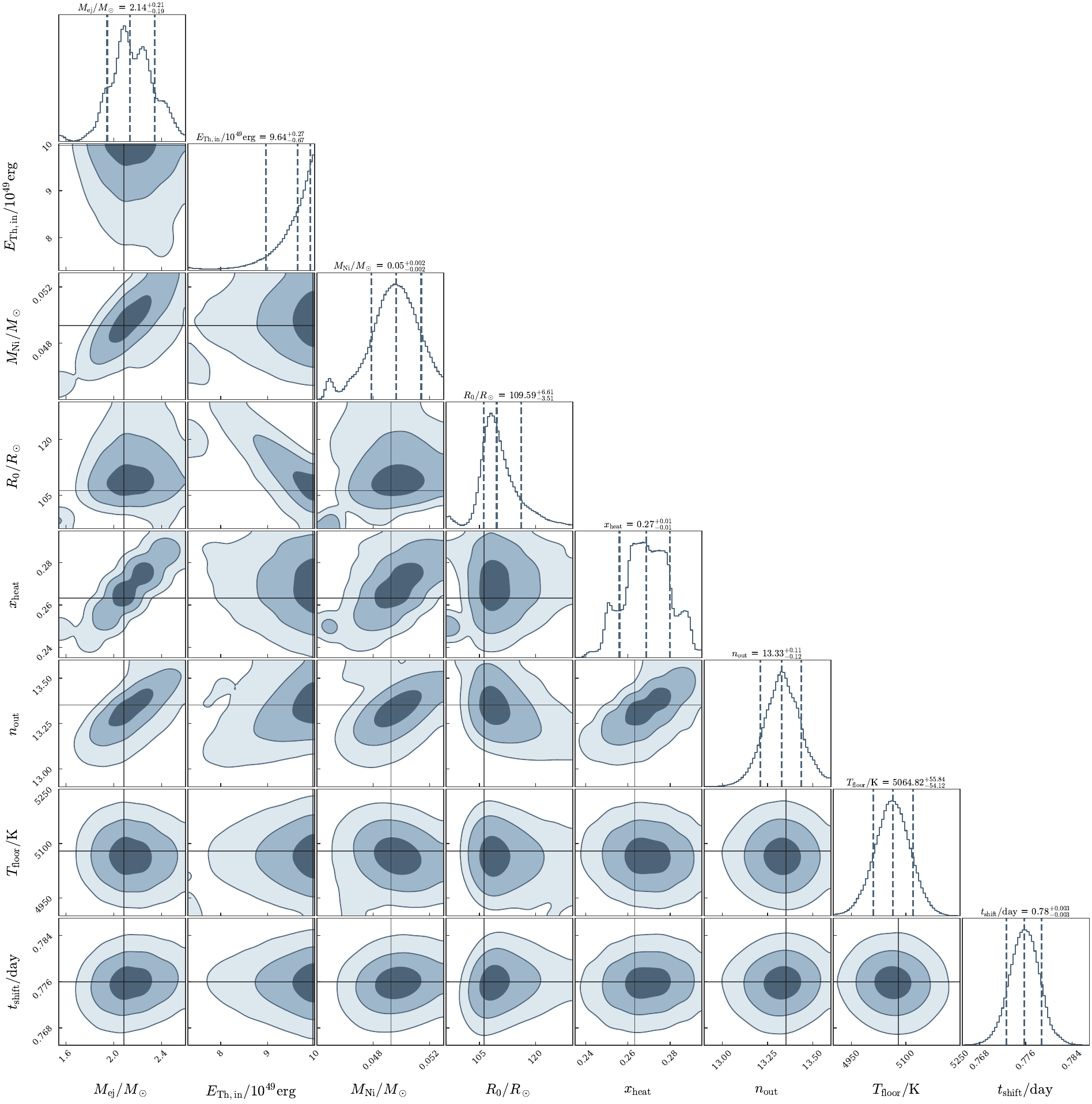}
    \caption{
    Corner plot of the posterior probability distributions of the physical parameters of SN~2024aecx obtained from the MCMC analysis with the extended \texttt{TransFit} model. 
    }
    \label{fig:corner}
\end{figure*}

Figure~\ref{fig:corner} shows the posterior distributions of the fitted
parameters. The main physical parameters are well localized, indicating
that the multiband light curves provide strong constraints on the ejecta
mass, nickel mass, effective outer radius, and outer density slope within
the adopted model. The steep inferred value of \(n_{\rm out}\) supports
a rapidly declining, low-mass outer envelope, while the low ejecta mass
is consistent with the rapid evolution of the radioactive peak.

\section{Discussion and Conclusions}
\label{sec:discussion}

SN~2024aecx provides a useful test case for double-peaked
Type~IIb light-curve modeling because its ultraviolet--optical
evolution combines three linked features: a short-lived early
cooling peak, a radioactive main peak at about 20~d, and a rapid
post-peak optical decline. In this work, we have modeled these
features with an extended version of \texttt{TransFit}, in which
the initially shock-deposited thermal energy and the subsequent
radioactive heating  are evolved through the same time-dependent diffusion
calculation. This treatment differs from two-component approaches
in which the shock-cooling and radioactive phases are added as
separate light-curve components. It therefore allows the early
cooling peak, the main radioactive peak, and the late optical
decline to be interpreted as coupled constraints on the same
ejecta structure.

The early cooling peak favors an extended but low-mass outer structure
around a strongly stripped progenitor. Motivated by the standard Type~IIb
picture, in which a compact helium-rich core is surrounded by a dilute
hydrogen-rich envelope, we adopted a stratified ejecta
density profile and fitted the outer-envelope slope. The inferred steep
outer density gradient, $n_{\rm out}\simeq13.3$, indicates that little
mass remains at large radii. This naturally produces a narrow
shock-cooling peak and a rapid decline toward the valley, consistent with
analytic expectations for cooling emission from extended low-mass
material \citep{Piro2015,Sapir2017}. The fitted
effective radius, $R_0\simeq1.1\times10^2\,R_\odot$, should be understood
as the characteristic radius of the emitting outer material at the onset
of the cooling phase, rather than as a unique hydrostatic stellar radius.

The secondary peak is reproduced with a low ejecta mass and a modest
nickel mass, $M_{\rm ej}\simeq2.1\,M_\odot$ and
$M_{\rm Ni}\simeq0.05\,M_\odot$. These values are consistent with a
strongly stripped Type~IIb progenitor and a short effective diffusion
timescale. Compared with previous analyses of SN~2024aecx, our inferred
radius is smaller than the two analytic shock-cooling estimates of
\citet{Zou2026} and is larger than the value inferred from the
two-component model of \citet{Xi2026}. The differences are expected
because the present calculation does not add an independent
shock-cooling component to an independent radioactive component, but
instead evolves both energy reservoirs through the same ejecta structure.

The rapid decline after the secondary maximum is the main tension with
the standard radioactive-diffusion picture. Using the same physical
parameters that reproduce the early peak and the main peak, the control
model with $f_{\rm out}=1$ fades too slowly, even after including
standard $\gamma$-ray leakage. We therefore introduced the effective
optical-output factor $f_{\rm out}(t)$ to describe the additional
suppression of the ultraviolet--optical luminosity. This factor is
phenomenological and should not be interpreted as a unique microscopic
thermalization efficiency. Possible physical origins include enhanced
nonlocal energy escape, redistribution of optical radiation into the
infrared, dust formation or dust echo effects, ejecta asymmetry or
clumping, and time-dependent opacity or line-formation effects
\citep{Maeda2008,Jerkstrand2015,Gall2014,Dimitriadis2017}.
In particular, the near-infrared excess reported by
\citet{Tinyanont2026} provides an observational motivation for considering
infrared redistribution or dust-related effects in SN~2024aecx, although
our light-curve modeling alone cannot determine whether this process is
responsible for the rapid optical fading.

Several limitations should be kept in mind. The model assumes spherical
symmetry, a gray optical opacity, and an approximate treatment of the
emergent spectral energy distribution. The adopted density profile is a
compact representation of Type~IIb ejecta rather than a unique
hydrodynamic reconstruction of the progenitor. The late-time factor
$f_{\rm out}(t)$ only quantifies the missing optical luminosity and
cannot by itself distinguish among infrared redistribution, dust effects,
geometry, opacity evolution, or nonlocal energy escape. Late-time
spectroscopy, infrared photometry, and detailed radiation-transfer
calculations are needed to identify the dominant physical mechanism.

In summary, SN~2024aecx supports the standard interpretation that
double-peaked Type~IIb light curves are produced by early shock cooling
followed by radioactive heating. The early light curve favors a steep,
low-mass extended envelope, while the main peak requires a low ejecta
mass and a modest nickel mass. However, the fast optical fading after the
secondary peak cannot be reproduced by standard radioactive diffusion
with $\gamma$-ray leakage alone. The late-time evolution therefore
contains additional information about energy escape, radiation
redistribution, and ejecta structure beyond the simplest
radioactive-diffusion model.

\begin{acknowledgements}
This work is supported by the National Key R\&D Program of China
(2021YFA0718500), the National Natural Science Foundation of China
(grant Nos. 12303047 and 12393811), the Natural Science Foundation of
Hubei Province (2023AFB321), and the China Manned Space Project
(CMS-CSST-2021-A12).
\end{acknowledgements}

\bibliography{sample701}{}
\bibliographystyle{aasjournalv7}

\end{document}